\documentclass{article}

\usepackage{arxiv}

\usepackage[utf8]{inputenc} 
\usepackage[T1]{fontenc}    
\usepackage{hyperref}       
\usepackage{url}            
\usepackage{booktabs}       
\usepackage{amsfonts}       
\usepackage{nicefrac}       
\usepackage{microtype}      
\usepackage{lipsum}		
\usepackage{graphicx}
\usepackage{natbib}
\usepackage{doi}

\title{STCALIR: Semi-Synthetic Test Collection for Algerian Legal Information Retrieval}

\author{ \href{https://orcid.org/0000-0003-4965-6185}{\includegraphics[scale=0.06]{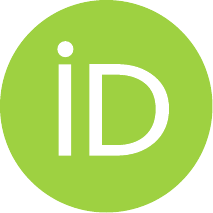}\hspace{1mm}M'hamed Amine Hatem} \\
	Laboratoire LITAN, École supérieure en Sciences \\
    et Technologies de l’Informatique et du Numérique\\
    RN 75, Amizour 06300, Bejaia, Algérie\\
    Bejaia, Algeria \\
	\texttt{hatem@estin.dz} \\
	\And
	\href{https://orcid.org/0009-0000-9632-2423}{\includegraphics[scale=0.06]{orcid.pdf}\hspace{1mm}Sofiane Batata} \\
	Ecole nationale Supérieure d’Informatique ESI \\
    Oued Smar Alger, Algérie\\
    Alger, Algeria\\
	\And
    \href{https://orcid.org/0009-0008-4835-041X}{\includegraphics[scale=0.06]{orcid.pdf}\hspace{1mm}Amine MAMMASSE} \\
	Laboratoire LITAN, École supérieure en Sciences \\
    et Technologies de l’Informatique et du Numérique\\
    RN 75, Amizour 06300, Bejaia, Algérie\\
    Bejaia, Algeria \\
	\And
    \href{https://orcid.org/0000-0002-2508-9158}{\includegraphics[scale=0.06]{orcid.pdf}\hspace{1mm}Faiçal Azouaou} \\
	Laboratoire LITAN, École supérieure en Sciences \\
    et Technologies de l’Informatique et du Numérique\\
    RN 75, Amizour 06300, Bejaia, Algérie\\
    Bejaia, Algeria \\
}

\renewcommand{\shorttitle}{STCALIR: Semi-Synthetic Test Collection for Algerian Legal Information Retrieval}

\hypersetup{
pdftitle={A template for the arxiv style},
pdfsubject={q-bio.NC, q-bio.QM},
pdfauthor={David S.~Hippocampus, Elias D.~Striatum},
pdfkeywords={First keyword, Second keyword, More},
}

\begin{document}
\maketitle

\begin{abstract}
Test collections are essential for evaluating retrieval and re-ranking models. However, constructing such collections is challenging due to the high cost of manual annotation, particularly in specialized domains like Algerian legal texts, where high-quality corpora and relevance judgments are scarce. To address this limitation, we propose STCALIR, a framework for generating semi-synthetic test collections directly from raw legal documents. The pipeline follows the Cranfield paradigm, maintaining its core components of topics, corpus, and relevance judgments, while significantly reducing manual effort through automated multi-stage retrieval and filtering, achieving a 99\% reduction in annotation workload. We validate STCALIR using the Mr. TyDi benchmark, demonstrating that the resulting semi-synthetic relevance judgments yield retrieval effectiveness comparable to human-annotated evaluations (Hit@10 $\approx$ 0.785). Furthermore, system-level rankings derived from these labels exhibit strong concordance with human-based evaluations, as measured by Kendall’s $\tau$ (0.89) and Spearman’s $\rho$ (0.92). Overall, STCALIR offers a reproducible and cost-efficient solution for constructing reliable test collections in low-resource legal domains.
\end{abstract}

\keywords{test collection \and synthetic \and Arabic legal text \and information retrieval}

\section{Introduction}
Information retrieval (IR) systems depend critically on high-quality test collections to evaluate performance and guide model development. These test collections generally comprise three components: a document corpus, a set of representative queries (topics), and corresponding relevance judgments, following the Cranfield paradigm, as outlined by Ellen M. Voorhees \cite{Voorhees2019} . In specialized domains, such as legal texts, constructing these collections is particularly challenging. The process is resource-intensive, time-consuming, and often requires domain expertise to manually assess the relevance of documents for each query. This challenge is further amplified in low-resource settings, such as the Algerian legal domain, which currently lacks any publicly available annotated corpora.
Recent efforts in IR have explored the use of synthetic test collections generated by large language models to reduce annotation costs \cite{10.1145/3626772.3657942,10.1145/3701716.3715311, Trkmen2025GenTRECTF}. However, many of these approaches rely on proprietary models, making them expensive and less accessible for low-resource domains.
To address these limitations, we introduce STCALIR (Sem-Synthetic Test Collection for Algerian Legal Information Retrieval), which serves two complementary roles in this study. First, it denotes the proposed framework used to generate semi-synthetic test collections from raw legal texts. Second, it refers to the semi-synthetic Arabic legal test collection produced using this framework. For clarity, we use STCALIR Framework when discussing the methodology and STCALIR Dataset when referring to the resulting test collection. The proposed methodology incorporates human-generated queries and leverages multiple retrieval systems—such as BM25 \cite{10.1145/3726302.3729905} and a set of bi-encoder models \cite{reimers-gurevych-2019-sentence} to create a pooled set of candidate passages at the first stage. In the second stage, a set of cross-encoder models \cite{10.1145/3404835.3462812, nogueira2020passagererankingbert} evaluates relevance to produce the top 10 candidates, which are subsequently assessed by human experts to generate the final relevance judgments. 

Our methodology preserves the multi-system pooling strategy of traditional test collections while drastically reducing human effort by 99\%. We validate the STCALIR Framework on the Mr. TyDi datasets \cite{zhang-etal-2021-mr}, demonstrating that synthetic relevance judgments approximate human-labeled collections. 

In addition to STCALIR Framework and STCALIR Dataset, we share all artifacts of this work\footnote{\begin{minipage}[t]{0.9\linewidth}
\raggedright Artifacts are available at (\url{https://huggingface.co/hatemestinbejaia})\end{minipage}}, along with our code\footnote{\begin{minipage}[t]{0.9\linewidth}
\raggedright Code is available at (\url{https://github.com/hatemamine/STCALIR-Semi-Synthetic-Test-Collection-for-Algerian-Legal-IR})\end{minipage}}, to benefit the community and support reproducible research in Arabic legal information retrieval.
\section{Related Work}
\subsection{Fully Synthetic Test Collections}
Hossein A. Rahmani et al.\cite{10.1145/3626772.3657942,10.1145/3701716.3715311} investigate constructing fully synthetic test collections using LLMs to reduce the cost and effort of traditional Cranfield style evaluation. By generating both queries and relevance judgments, their workflows achieve high system-ranking correlations with human-annotated benchmarks (Kendall’s $\tau \approx 0.857$), demonstrating that synthetic collections can reliably preserve relative system ordering. Despite being fully synthetic, low-quality queries still required filtering by human assessors, highlighting the need for human-in-the-loop oversight. Similarly, Mehmet Deniz Türkmen et al. \cite{Trkmen2025GenTRECTF} propose generating both documents and relevance labels entirely with LLMs, creating a 96,196-document corpus from 300 topics at a minimal cost (\$126), showing the potential for low-cost, fully synthetic evaluation resources.

\subsection{Synthetic Relevance Labeling and Test Collection Augmentation}
Björn Engelmann et al. \cite{10.1145/3726302.3730342} introduce REANIMATOR, a framework for automatically assigning synthetic relevance labels to content extracted from unstructured PDF allowing the extension and repurposing of existing test collections for multiple IR tasks such as document, passage, and table retrieval or retrieval-augmented generation (RAG). REANIMATOR demonstrates how synthetic labeling can enrich datasets like TREC-COVID, producing a multimodal evaluation resource while reducing reliance on new human annotations. Rahmani et al. \cite{10.1145/3746252.3760908} examine biases in LLM-generated IR collections, finding that while synthetic queries tend to overestimate absolute system performance, relative ranking correlations remain high (Kendall’s $\tau$ = 0.72–0.91), and GPT-based evaluators exhibit a self-preference bias, reinforcing the need for careful validation of synthetic labels.

\subsection{Synthetic Query Generation and Data Fusion}
Timo Breuer \cite{10.1145/3673791.3698423} and Alaofi et al. \cite{10.1145/3539618.3591960} explore the use of instruction-tuned LLMs to generate multiple query variants for retrieval and test collection expansion. Breuer demonstrates a four-step pipeline—prompting, query generation, retrieval, and Reciprocal Rank Fusion (RRF)—finding that including topic titles, descriptions, and narratives consistently improves performance across TREC newswire benchmarks and uncovers previously unjudged relevant documents. Alaofi et al. show that GPT-3.5-generated query variants, although less diverse than human queries, can achieve substantial overlap with human-generated pools (up to 71.1\% for relevant documents at depth 100), highlighting the potential of LLM-generated variants for augmenting document pooling in test collection construction.

\subsection{Domain-Adaptive Synthetic Data for Sensitive Environments}
Helia Hashemi et al. \cite{10.1145/3578337.3605127,vakili2025data} propose frameworks for domain-adaptive dense retrieval and clinical NLP tasks using synthetic corpora. By leveraging target-domain descriptions, LLMs generate synthetic documents, domain-aligned queries, and pseudo-relevance labels, with iterative retrieval and teacher-student labeling ensuring relevance. Experiments demonstrate that dense retrievers trained on these synthetic datasets achieve strong performance without accessing the original target collection, and in clinical settings, moderately-sized LLMs can produce high-utility machine-annotated corpora for NER tasks with minimal performance loss. Success depends primarily on the quality of the annotating model and seed documents rather than LLM size, showing that cost-effective, privacy-preserving synthetic data generation is feasible, albeit with limitations for more complex tasks and full privacy assurance.
\section{Methodology}
\begin{figure}
	\centerline{\includegraphics[width=0.8\linewidth]{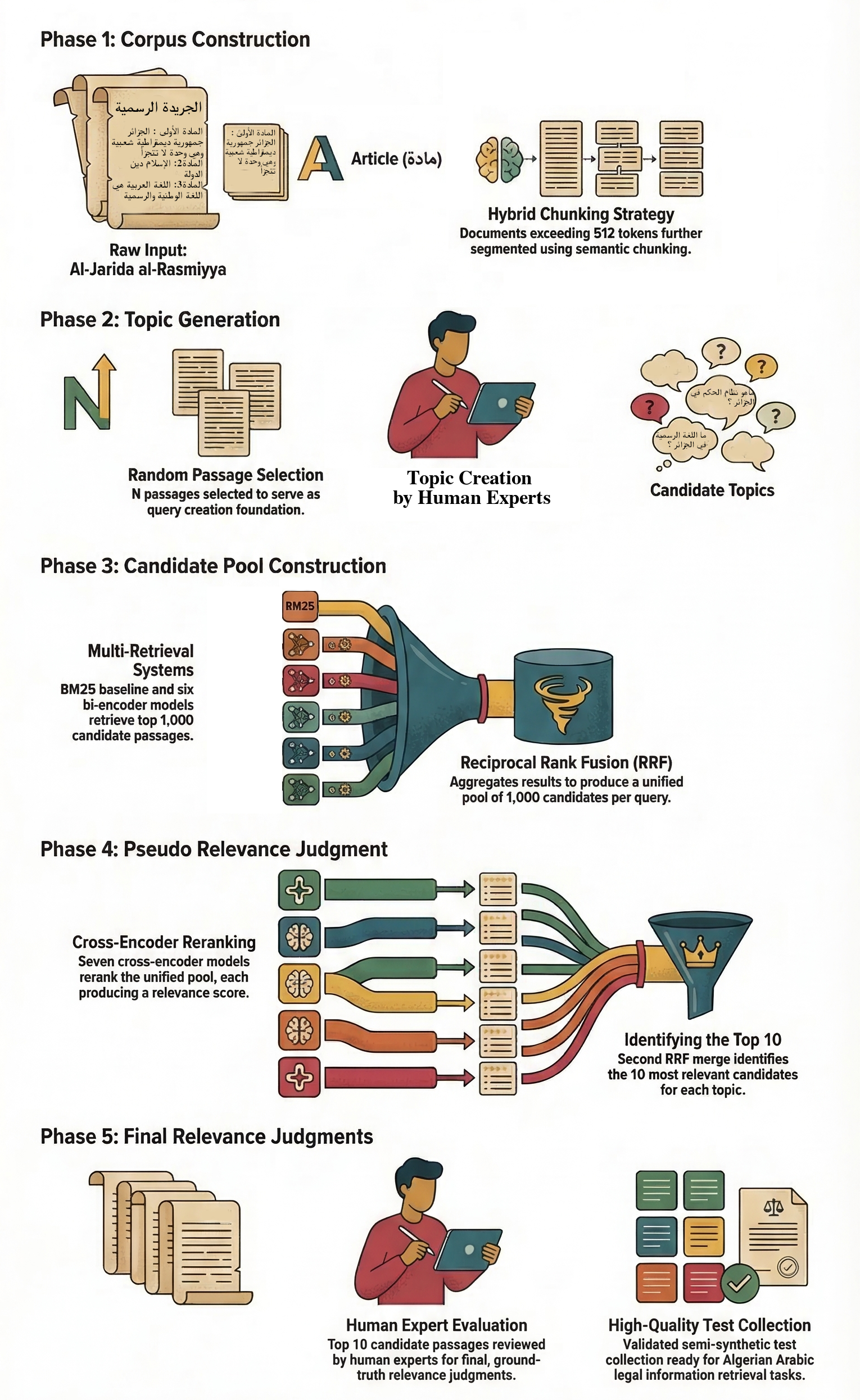}}
	\caption{5 phase methodology for building semi-synthetic test collection}
	\label{methodology}
\end{figure} 
In contrast to recent synthetic test-collection pipelines that depend on expensive proprietary LLMs for both topic generation and relevance assessment, the STCALIR framework is fully based on open-source models that can be executed on a single GPU, rendering the approach cost-efficient and fully reproducible. Moreover, STCALIR incorporates human experts in the final relevance judgment stage, enhancing the validity and reliability of the resulting test collection, while simultaneously reducing the required human annotation effort by approximately 99\% compared with traditional fully manual test collections.
The core idea behind STCALIR relies on the observed capabilities of retrieval models \cite{10.1145/3796229}, which were fine-tuned on mMARCO \cite{bonifacio2022mmarcomultilingualversionms} and evaluated in a zero-shot scenario on the Mr. TyDi dataset. The bi-encoder models achieve an average Recall@1000 of $\approx 0.7$, indicating that the initial retrieval stage captures approximately 70\% of relevant documents per topic. By merging multiple retrieval outputs via Reciprocal Rank Fusion (RRF) \cite{10.1145/1571941.1572114}, coverage can be further increased to around 80\%, forming a robust candidate pool for re-ranking. In the second stage, cross-encoder models—achieving an MRR@10 of $\approx 0.24$ in the zero-shot evaluation—refine the ranking to elevate the most relevant documents to the top positions, ensuring that, on average, at least two truly relevant documents appear in the top 10. This two-stage pipeline leverages the complementary strengths of high-recall bi-encoders and high-precision cross-encoders, substantially reducing the human effort required for final relevance annotation, while producing a scalable and reproducible benchmark for Arabic legal information retrieval.
STCALIR Framework is structured as a five-phase pipeline (see Figure \ref{methodology}) designed to construct a semi-synthetic test collection for Arabic legal information retrieval. A standard test collection consists of three components: a document corpus, a set of topics, and a corresponding set of relevance judgments. Each phase of STCALIR Framework addresses a specific aspect of the test-collection creation process, integrating automated methods with minimal human validation to ensure both efficiency and quality.
\subsection{phase 1 : Corpus Acquisition and Preprocessing:} 
\begin{figure}
	\centerline{\includegraphics[width=\linewidth]{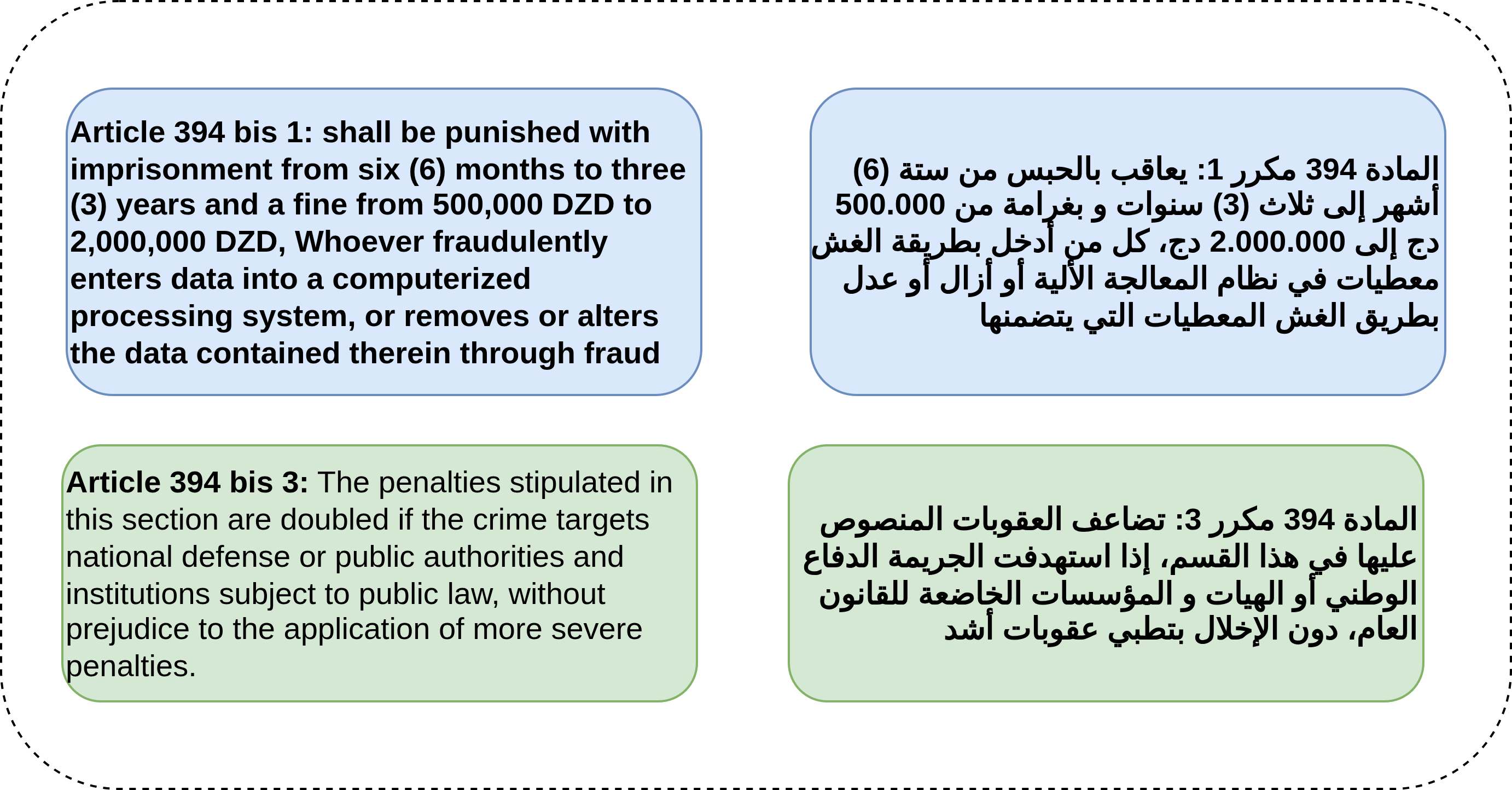}}
	\caption{Side-by-Side Display of Original Algerian Arabic Legal Text and Its English Translation from the Algerian Official Gazette (Issue 71, 10 November 2004, Page 12)}
	\label{stcalirEX}
\end{figure}

The raw Algerian legal texts was obtained from the publicly accessible legislative repository of the Official Gazette of the People's Democratic Republic of Algeria \footnote{(\url{https://www.joradp.dz})}, which provides authenticated legal documents exclusively in PDF format. Due to variations in document structure, layout, and typography—particularly in older issues automated text extraction required optical character recognition (OCR) \cite{10.1145/3768150, alghyaline2023arabic} to generate machine-readable content, which underwent a thorough post-processing stage, during which inconsistencies introduced by OCR noise were manually reviewed, corrected, and validated by human experts.
Given that legal documents often vary substantially in length—frequently exceeding the maximum token limits imposed by contemporary information retrieval (IR) models—we adopt a hybrid chunking strategy \cite{GomezCabello2025ComparativeEO} to divide the text into manageable units while preserving contextual and semantic continuity. Each resulting segment maintains both the surrounding context and the internal semantic coherence required for reliable downstream retrieval tasks.
Because legal texts are typically organized into articles introduced by explicit lexical markers as illustrated in Figure \ref{stcalirEX}. We first perform structural chunking using the pattern “article”. This allows us to align segmentation with the natural hierarchy of legal documents. However, when an individual article still surpasses the input size constraints of specific IR models, we employ semantic chunking as a secondary method.
Semantic chunking partitions the text based on meaning rather than formal structure, producing conceptually coherent segments. This approach relies on embedding-based similarity measures to identify logical semantic boundaries. While it improves intra-chunk semantic integrity, it also introduces higher computational overhead due to the deeper linguistic processing required.
The final corpus produced in Phase 1 comprises 105,201 passages, with their key characteristics summarized in Table \ref{tab:dataset_subsets}.
\subsection{Phase 2 : Topic creation:} 
\begin{figure}
	\centerline{\includegraphics[width=\linewidth]{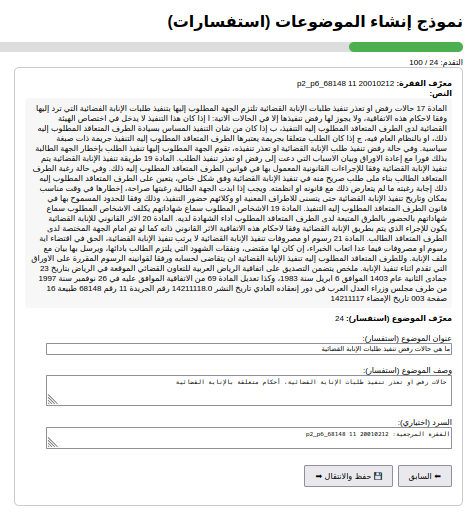}}
	\caption{Screenshot showing the Web Topics annotation interface}
	\label{TopicsAnnotation}
\end{figure}

Human expertise is employed for query creation. To facilitate this process, a dedicated web interface is used as illustrated in Figure \ref{TopicsAnnotation}, ensuring efficient and consistent topic formulation while maintaining comprehensive coverage of the information needs reflected in the underlying corpus.

\subsection{Phase 3: Candidate Pool Construction:} A set of retrieval systems—including a BM25 baseline and fine-tuned bi-encoder models  retrieve the top 1,000 candidate passages per topic. Outputs are merged using Reciprocal Rank Fusion (RRF) to increase coverage and reduce the likelihood of missing relevant documents.

\subsection{Phase 4: Pseudo Relevance Judgement:} a set of Cross-encoder models evaluate the merged candidate pool to estimate relevance scores. These scores are again fused via RRF to identify the top 10 passages per topic, prioritizing documents with the highest relevance likelihood based on fine-grained semantic understanding.

\subsection{Phase 5: Human Validation of Relevance Judgments:} 
\begin{figure}
	\centerline{\includegraphics[width=\linewidth]{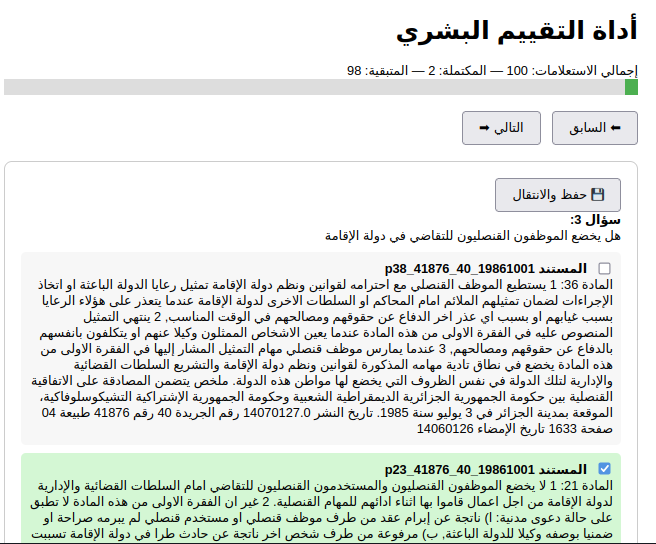}}
	\caption{Screenshot showing the Web Interface for human relevance assessment}
	\label{humanAnnotationForm}
\end{figure}

The top 10 ranked candidates undergo final human review, ensuring that the resulting relevance judgments are both accurate and reliable while maintaining the integrity of the test collection. This process is facilitated through a dedicated web interface, as illustrated in Figure \ref{humanAnnotationForm}.
This five-phase pipeline leverages the complementary strengths of high-recall bi-encoders and high-precision cross-encoders, producing a scalable, reproducible, and cost-efficient benchmark suitable for evaluating retrieval and re-ranking models in low-resource legal domains.
To assess the effectiveness of the proposed methodology, we performed experiments on the widely used Mr. TyDi benchmark dataset (see Table \ref{tab:dataset_subsets}). The pipeline was applied to produce synthetic relevance judgments derived from the Mr. TyDi corpus and corresponding topic set.
\begin{table*}[htbp]
\centering
\caption{Detailed summary of the dataset  used in this study, including their sizes and feature characteristics.}
\label{tab:dataset_subsets}
\begin{tabular*}{\linewidth}{@{\extracolsep{\fill}} p{5cm} p{4cm} p{4cm} }
\toprule
\textbf{Test collection} & \textbf{Size} & \textbf{Features }  \\
\midrule
\textbf{Mr. TyDi test collection Arabic} &
The Arabic corpus contains 2,106,586 passages, Number of Queries (Topics): Test: 1,081 queries, Number of Judgments: Test: 1,257 judgments &
\textbf{Passage length:} 59.50±73.76 words (\textit{empirically measured}), with lengths ranging from 0 to 9317 words. \newline
\textbf{Query length:} 6.63±2.13 words (\textit{empirically measured}) with lengths ranging from 3 to 18 words \newline
\textbf{language: } Arabic  \\
\midrule
\textbf{STCALIR } &
STCALIR corpus contains 105,201 passages, Number of Queries (Topics): 100 queries, Number of Judgments: 1,000 judgments (Number of Relevant Judgments: 434)  &
\textbf{Passage length:} 85.42 ± 62.04 words (\textit{empirically measured}), with lengths ranging from 20 to 447 words \newline 
\textbf{Query length:} 7.11 ± 3.08 words (\textit{empirically measured}), with lengths ranging from 2 to 16 words \newline 
\textbf{language: } Arabic \\
\bottomrule
\end{tabular*}
\end{table*}

\section{Experiment}
To assess the reliability and effectiveness of the STCALIR Framework, we conducted a series of experiments using both the STCALIR Dataset and the Mr. TyDi benchmark dataset. 
To select the most suitable retrieval and re-ranking models, we evaluated both in-domain performance and cross-domain generalization \cite{10.1007/978-3-031-99991-8_7, thakur2021beirheterogenousbenchmarkzeroshot}. Generalization was quantified using the Generalization Ratio (GR), defined as the zero-shot performance divided by the in-domain performance for a given metric (e.g., MRR@10), with values near or above 1 indicating robust performance on unseen datasets. From seven cross-encoder candidate models, five were retained for evaluation, as two exhibited poor zero-shot performance despite strong in-domain results. Model selection prioritized systems that combined high in-domain metrics with strong GR values, ensuring effectiveness across diverse datasets rather than overfitting to a specific corpus.
The evaluation followed the five-phase structure of the proposed pipeline. For the STCALIR Dataset, the corpus and topics were generated in Phases 1 and 2, whereas for Mr. TyDi, the existing corpus, topics, and relevance judgments were directly loaded. For both datasets, Phases 3, 4, and 5 were applied as follows:
Phase 3 — Initial Candidate Retrieval.
Using the corpus and topics produced in Phase 1, we first applied the Pyserini library \cite{10.1145/3404835.3463238} to retrieve the top 1,000 candidate passages per topic using the BM25 retrieval model. In parallel, six bi-encoder models \footnote{%
\begin{minipage}[t]{0.9\linewidth} 
\raggedright
\url{https://huggingface.co/hatemestinbejaia/mmarco-Arabic-AraElectra-bi-encoder-KD-v1} \\
\url{https://huggingface.co/hatemestinbejaia/mmarco-Arabic-AraElectra-bi-encoder-NoKD-v1} \\
\url{https://huggingface.co/hatemestinbejaia/mmarco-Arabic-AraDPR-bi-encoder-KD-v1} \\
\url{https://huggingface.co/hatemestinbejaia/mmarco-Arabic-AraDPR-bi-encoder-NoKD-v1} \\
\url{https://huggingface.co/hatemestinbejaia/mmarco-Arabic-mMiniLML-bi-encoder-KD-v1} \\
\url{https://huggingface.co/hatemestinbejaia/mmarco-Arabic-mMiniLML-bi-encoder-NoKD-v1}
\end{minipage}%
} were used to encode both corpus and topic representations, and dense retrieval was performed with FAISS \cite{11202651}, yielding 1,000 candidates from each bi-encoder. The resulting candidate lists were merged using Reciprocal Rank Fusion (RRF), producing a unified pool of 1,000 candidate passages per topic.

Phase 4 — Re-ranking with Cross-Encoders.
From this fused pool, five cross-encoder models \footnote{%
\begin{minipage}[t]{0.9\linewidth} 
\raggedright
\url{https://huggingface.co/hatemestinbejaia/mmarco-Arabic-AraElectra-cross-encoder-KD-v1} \\
\url{https://huggingface.co/hatemestinbejaia/mmarco-Arabic-AraElectra-cross-encoder-NoKD-v1} \\
\url{https://huggingface.co/hatemestinbejaia/mmarco-Arabic-AraDPR-cross-encoder-KD-v1} \\
\url{https://huggingface.co/hatemestinbejaia/mmarco-Arabic-AraDPR-cross-encoder-NoKD-v1} \\
\url{https://huggingface.co/cross-encoder/mmarco-mMiniLMv2-L12-H384-v1}
\end{minipage}%
} were applied to perform fine-grained re-ranking. Their outputs were then combined using Reciprocal Rank Fusion (RRF), from which the top-10 most relevant passages per topic were selected as the system’s predicted relevance labels.

Phase 5 — Human Validation.
Human annotators reviewed the top-10 ranked passages and assigned the final relevance judgments, ensuring high-quality and reliable ground-truth labels.

To assess the validity of the STCALIR Framework, we evaluated how closely the semi-synthetic relevance judgments align with the fully manual human annotations in the Mr. TyDi dataset. 
Specifically, we used Hit@10 to evaluate how well the system rankings derived from the semi-synthetic judgments recover the top-ranked documents identified by the original human-labeled assessments. Hit@10 measures the proportion of queries for which at least one relevant document appears among the top 10 retrieved results, making it a practical indicator of retrieval effectiveness at the system level. Higher Hit@10 values therefore indicate stronger alignment with expert annotations, providing a robust measure of the reliability and validity of the STCALIR-generated test collection.
The experiments were conducted in a \textbf{Kaggle Tesla P100-PCIE 16GB} environment. The framework does not require specialized hardware or proprietary software, relying solely on open-source libraries. This design ensures that the complete workflow is fully reproducible without incurring additional hardware or software costs, making it accessible to researchers operating in low-resource computational settings.
\section{Results and Discussion}
The evaluation of the STCALIR Framework focuses on two key aspects: (1) the reliability of the semi-synthetic relevance judgments generated by the pipeline, as measured by system ranking consistency, and (2) the efficiency gains achieved through the proposed multi-stage pooling and filtering strategy. All experiments were conducted on the Mr. TyDi benchmark dataset, which provides a well-established reference for assessing ranking performance. Models used in the pipeline were fine-tuned on the MS MARCO dataset and evaluated in a zero-shot transfer setting on Mr. TyDi.
\subsection{Ranking Reliability}
To measure the effectiveness of the semi-synthetic relevance judgments, we compared system rankings produced by the STCALIR pipeline with the original human-labeled Mr. TyDi relevance assessments. Hit@10 was used to quantify ranking agreement. Across retrieval and re-ranking configurations, the semi-synthetic judgments successfully retrieved at least one relevant document in the top 10 for approximately 78\% of queries (Hit@10 $ \approx $ 0.78), indicating strong alignment with human annotations, indicating that the semi-synthetic labels closely approximate the system-level ordering induced by fully human-annotated judgments. 
\subsection{Effectiveness of Multi-Stage Pooling}
The candidate retrieval strategy in STCALIR is designed to ensure high recall while maintaining strong ranking quality in the final annotated subset. In the initial stage, dense bi-encoder models are employed in a zero-shot setting to retrieve candidate passages. Empirically, these models achieve high recall, enabling the system to capture a substantial portion of relevant documents within the top-ranked candidates.
To further improve coverage, the outputs of multiple bi-encoders are combined using Reciprocal Rank Fusion (RRF). As shown in Table \ref{tab:impact_pipline}, this multi-system pooling strategy improves top-10 effectiveness compared to individual retrieval models. While BM25 alone achieves a Hit@10 of 0.569, combining multiple retrieval systems increases this value to 0.654, demonstrating the benefit of aggregating diverse retrieval signals to enhance candidate quality.
In the second stage, cross-encoder models are applied to re-rank the pooled candidates. This step significantly improves the quality of the top-ranked results. Using a single cross-encoder already leads to substantial gains, increasing Hit@10 to 0.766 and MRR to 0.585. Further improvements are observed when combining multiple cross-encoders via RRF, with the full pipeline achieving the best performance (Hit@10 = 0.785, nDCG@10 = 0.656). These results indicate that cross-encoder fusion enhances the precision of top-ranked documents by leveraging complementary modeling capabilities.
Interestingly, increasing the number of cross-encoders beyond three yields only marginal improvements, suggesting diminishing returns from additional models. This observation highlights that a relatively small ensemble of cross-encoders is sufficient to achieve near-optimal performance, providing a favorable balance between effectiveness and computational cost.
Overall, the results in Table \ref{tab:impact_pipline} demonstrate that the proposed multi-stage pipeline effectively combines high-recall retrieval with high-precision re-ranking. This leads to a top-10 candidate set that contains a high proportion of relevant documents, thereby enabling efficient human annotation while preserving evaluation quality.
\begin{table}[h]
\centering
\caption{Impact of pooling and re-ranking strategies on top-10 retrieval effectiveness.}
\label{tab:impact_pipline}
\begin{tabular}{l c c c}
\hline
\textbf{Setting} & \textbf{Hit@10} & \textbf{MRR} & \textbf{nDCG@10} \\
\hline
BM25 only pooling & 0.569 & 0.353 & 0.444 \\
Best single bi-encoder (mMiniLM KD) & 0.496 & 0.303 & 0.384 \\
BM25 + multi bi-encoder (RRF) & 0.654 & 0.449 & 0.541 \\
\hline
+ Best single cross-encoder \\ mmarco-mMiniLMv2-L12-H384-v1 \footnote{\url{https://huggingface.co/cross-encoder/mmarco-mMiniLMv2-L12-H384-v1}} & 0.766 & 0.585 & 0.655 \\
+ 3 cross-encoders +RRF (full pipeline)& \textbf{0.785} & \textbf{0.585} & \textbf{0.656} \\
+ 5 cross-encoders +RRF (full pipeline) & 0.785 & 0.581 & 0.654 \\
\hline
\end{tabular}
\end{table}

\subsection{Reduction in Human Annotation Effort}
\begin{figure}
	\centerline{\includegraphics[width=\linewidth]{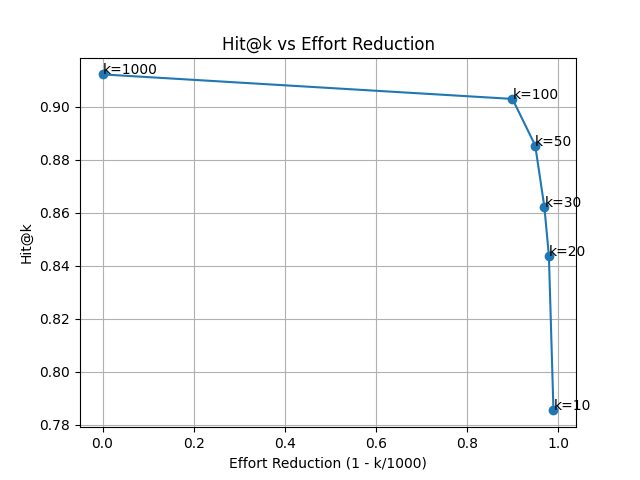}}
	\caption{Screenshot showing the Web Topics annotation interface}
	\label{Hit@k_vs_effort_reduction}
\end{figure}

A key objective of STCALIR is to significantly reduce the cost of manual relevance annotation while preserving evaluation quality. By restricting human assessment to only the top-10 retrieved passages per query, the proposed framework achieves up to a 99\% reduction in annotation effort compared to traditional pooling-based approaches, which typically require evaluating up to 1000 candidate documents per topic.
To analyze the trade-off between annotation effort and retrieval effectiveness, we evaluate Hit@k at different cutoff levels and relate it to the corresponding effort reduction, defined as $1 - \frac{k}{1000}$. The results are illustrated in Figure \ref{Hit@k_vs_effort_reduction}. As the candidate set is progressively reduced, retrieval effectiveness decreases gradually rather than abruptly. Specifically, while the full candidate pool (k = 1000) achieves a Hit@k of 0.912, reducing the pool to k = 10—corresponding to a 99\% reduction in annotation effort—still retains a relatively high Hit@10 of 0.785.
This behavior indicates that the majority of relevant documents are concentrated within the top-ranked results, and that aggressive pruning of the candidate set has a limited impact on effectiveness. Even at intermediate cutoffs (e.g., k = 50 or k = 30), the reduction in performance remains modest, further supporting the robustness of the ranking pipeline.
Overall, these findings demonstrate that STCALIR achieves a favorable balance between efficiency and effectiveness. By focusing human annotation on a small, high-quality subset of candidates, the framework substantially reduces annotation costs while maintaining a strong approximation of full-pool relevance judgments.
\subsection{System-Level Ranking Correlation}
\begin{figure}
	\centerline{\includegraphics[width=\linewidth]{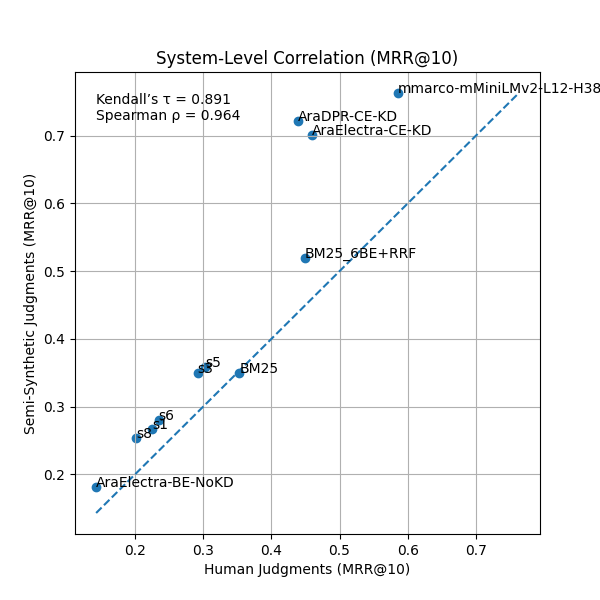}}
	\caption{System-level correlation between human and semi-synthetic relevance judgments for MRR@10. Each point represents one of the 11 retrieval systems. The dashed diagonal line indicates perfect agreement. The strong alignment confirms that the proposed framework preserves system ranking consistency}
	\label{systemLevelCorrelation}
\end{figure}

To assess whether the semi-synthetic relevance judgments preserve the relative ordering of retrieval systems, we compute system-level ranking correlation between rankings induced by human and semi-synthetic labels.  Table \ref{tab:correlation} reports Kendall’s $\tau$ and Spearman’s $\rho$ based on rankings derived from MRR@10, nDCG@10, and Recall@10.
Correlation is computed over 11 diverse retrieval systems, including sparse, dense, and hybrid approaches.The inclusion of multiple and diverse retrieval systems ensures that the correlation analysis is robust and not biased toward a specific model family.
The results show a high level of agreement between system rankings obtained using human and semi-synthetic relevance judgments (see Figure~\ref{systemLevelCorrelation}). Kendall’s $\tau$ values range from 0.818 to 0.891, while Spearman’s $\rho$ ranges from 0.918 to 0.964 across all metrics. These values indicate strong to near-perfect correlation, demonstrating that the proposed framework reliably preserves the relative ordering of retrieval systems.
This is particularly important in information retrieval evaluation, where the primary goal of a test collection is to enable reliable comparison between systems rather than to perfectly reproduce absolute effectiveness scores.
\begin{table}[h]
\centering
\caption{System-level correlation between human and semi-synthetic relevance judgments over 11 retrieval systems.}
\label{tab:correlation}
\begin{tabular}{lcc}
\toprule
\textbf{Metric} & \textbf{Kendall’s $\tau$} & \textbf{Spearman $\rho$} \\
\midrule
MRR@10    & \textbf{0.891} & \textbf{0.964} \\
nDCG@10   & 0.855 & 0.945 \\
Recall@10 & 0.818 & 0.918 \\
\bottomrule
\end{tabular}
\end{table}
\subsection{Evaluation on the STCALIR Dataset}
To assess the effectiveness of the proposed framework in the target domain, we evaluate STCALIR using semi-synthetic relevance judgments derived from the constructed dataset. Table \ref{tab:eval_STCALIR_dataset} reports the performance of different retrieval configurations in terms of MRR@10 and Recall@10.
The results show that BM25 provides a strong baseline, achieving an MRR@10 of 0.450 and a Recall@10 of 0.620. Incorporating multi-system pooling via RRF leads to a substantial improvement in recall, increasing Recall@10 to 0.700, while maintaining comparable ranking quality (MRR@10 = 0.448). This indicates that pooling enhances the coverage of relevant documents without degrading the overall ranking effectiveness.
The full pipeline, which combines multi-system pooling with cross-encoder re-ranking, yields the best performance across both metrics. In particular, MRR@10 increases significantly to 0.720, while Recall@10 reaches 0.97. This substantial improvement demonstrates the effectiveness of the re-ranking stage in promoting relevant documents to the top positions, while preserving the high recall achieved during the pooling phase.
These results confirm that the proposed multi-stage framework generalizes effectively to the STCALIR dataset and is well-suited for domain-specific retrieval tasks. The combination of high recall and strong top-ranked precision further supports the suitability of the framework for efficient annotation, as most relevant documents are successfully captured within the top-10 results.
\begin{table}[h]
\centering
\caption{Evaluation on the STCALIR dataset using semi-synthetic relevance judgments.}
\label{tab:eval_STCALIR_dataset}
\begin{tabular}{l c c}
\hline
\textbf{Setting} & \textbf{MRR@10} & \textbf{Recall@10} \\
\hline
BM25 & 0.450 & 0.620 \\
BM25 + multi-system pooling (RRF) & 0.448 & 0.700 \\
Full pipeline (pooling + cross-encoder RRF) & \textbf{0.720} & \textbf{0.97} \\
\hline
\end{tabular}
\end{table}
\subsection{Practical Advantages and Limitations}
The framework’s reliance on open-source models ensures domain-adaptable fine-tuning, transparent behavior, and minimal hallucination risk—critical when handling sensitive legal texts. Additionally, all components can be executed on a single GPU, making the pipeline accessible to research groups with limited computational resources.
We note several limitations:
Corpus quality dependency: The quality of the extracted corpus is influenced by the performance of OCR tools. Errors such as misrecognized characters, segmentation issues, or missing diacritics may affect downstream retrieval. Future improvements in Arabic OCR \cite{wasfy2025qariocrhighfidelityarabictext} technologies are expected to reduce errors, lower corpus-construction costs, and improve overall quality and usability of the text.
Model dependency: The effectiveness of STCALIR relies on the performance and generalization ability of bi-encoder and cross-encoder models. These models can be further adapted to the legal domain via lightweight fine-tuning methods, such as Low-Rank Adaptation (LoRA) \cite{hu2022lora}, to improve relevance estimation and ranking accuracy.
\section{Conclusions}
This work introduces STCALIR, a scalable and cost-efficient framework for constructing semi-synthetic test collections in low-resource domains, with a particular focus on Algerian Arabic legal texts. By combining manual human topic generation, multi-system bi-encoder retrieval, RRF-based pooling, and cross-encoder reranking, the pipeline effectively concentrates human annotation on a reduced set of high-value candidates, reducing manual effort by 99\% while preserving the core Cranfield principles.

Validation on the Mr. TyDi benchmark demonstrates that the resulting collection achieves high top-10 agreement (Hit@10 $\approx 0.78$) with human-annotated resources. Furthermore, system-level rankings derived from these labels exhibit strong concordance with human-based evaluations, as measured by Kendall’s $\tau$ (0.89) and Spearman’s $\rho$ (0.92), confirming the reliability of the semi-synthetic judgments.

Beyond its empirical performance, the STCALIR framework offers practical advantages: it operates entirely on open-source models that can be fine-tuned for domain specificity, avoids reliance on large proprietary LLMs, and minimizes hallucination risks in sensitive legal contexts.

\bibliographystyle{unsrtnat}
\bibliography{ref}  






\end{document}